\documentclass[12pt]{article}
\usepackage{graphicx}
\input{epsf}
\begin{document}

\title{Exclusive Diffraction at HERA and Beyond}
\author{ Salvatore Fazio\thanks{E-mail: fazio@fis.unical.it}\\
{\small Dipartimento di Fisica, Universit\'a della
Calabria and INFN-Cosenza,}\\ 
{\small I-87036 Arcavacata di Rende, Cosenza, Italy}
%Author3\\
%{\small Address3-}
}
\date{}
\maketitle
\begin{abstract}
The exclusive diffractive production of vector mesons and real photons in $ep$ collisions has been studied at HERA in a wide kinematic range. Here the most recent experimental results are presented together with a Regge-type model and projects for new diffractive studies at LHC.
\end{abstract}

\section{Introduction}

The diffractive scattering is a process where the colliding particles scatter at very small angles and without any color flux in the final state. This involves a propagator carrying the vacuum quantum numbers, called Pomeron and described, in the soft regime, within the Regge theory. Since the first operation period in 1992, ZEUS and H1, the two experiments dedicated to the DIS physics at HERA, observed that $(\sim 10 \%)$ of lepton-proton DIS events had a diffractive origin. It opens a new area of studies in diffractive production mechanism, providing a hard scale which can be
varied over a wide range and therefore it is an ideal testing for QCD models.
Moreover, the diffractive production of Vector Mesons (VMs) and real photons, allows to study the transition from the soft to the hard regime in strong interactions. The hard regime (high energy and low Bjorken-$x$, $x_{Bj}$) is sensitive to the gluon content and well described by perturbative-QCD, while in the soft regime (low-$x$) the interaction is well described within the Regge phenomenology. Indicating with $Q^2$ the virtuality of the exchanged photon and with $M^2$ the square mass of the produced VM, HERA data suggested a universal hard scale, $Q^2+M^2$, for the diffractive exclusive pruduction of VMs and real photons, which indicates the transition from the soft to the hard regime.
Moreover, the diffractive production of real photons, a process known as Deeply Virtual Compton Scattering (DVCS), leads to the extraction of the Generalized Parton Distribution functions (GPDs), containing combined imformations about the longitudinal momentum distribution of partons and their position on the trasfers plain. The GPD-based calculations will be very helpfull in the description of the Higgs boson diffractive production mechanism, which will be experimentally studied with the LHC accelerator. 

The following Sections will present the most recent results achieved at HERA together with a short outlook to the future exclusive diffraction program at LHC. An introduction to a new phenomenological model for the description of the VMs and DVCS amplitudes in the framework of the Regge theory will be also given.

\section{Exclusive diffraction at HERA}

\begin{figure}[t]
\centering
{\includegraphics[width=9.0cm]{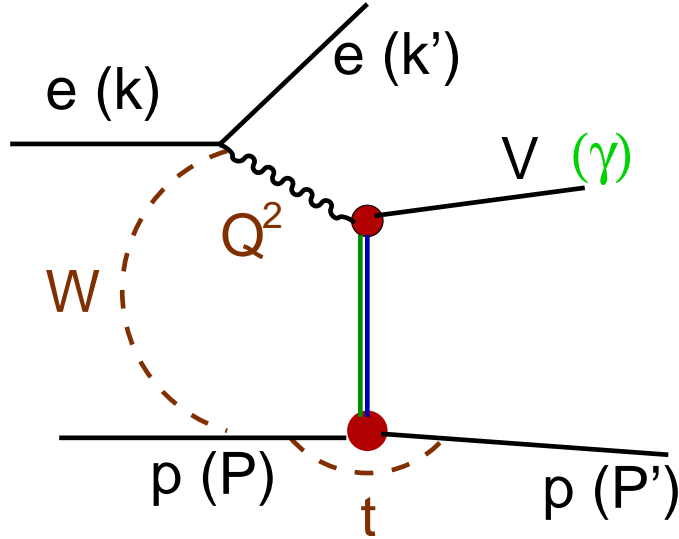}}
\caption {Typical Faynman diagram for the exclusive diffractive VM production at HERA, showing the relevant kinematic variables.}
\label{diff_diag}
\end{figure}

The diffractive processes are characterized by the presence of a leading proton
in the final state carrying most of the proton beam energy and by a large rapidity gap (LRG) in the forward (proton) direction.
Figure~\ref{diff_diag} shows a schematic diagram of the exclusive diffractive process at HERA, $ep\rightarrow -> eVp$, together with the relevant kinematic variables: the photon virtuality, $Q^2$, the photon-proton centre-of-mass energy, $W$ ,
and the square of the four-momentum transfer at the proton vertex, $t$. Relevant kinematic variables in diffraction are also the fraction
of the proton longitudinal momentum carried by the exchanged 
colour singlet object, $x_{IP}$, and the fraction of the exchanged momentum carried by the quark coupling to the virtual photon, $\beta$.

\subsection{The $Q^2$ and $W$ dependence of the cross section}

\begin{figure}[h]
\centering
{\includegraphics[width=0.7\textwidth]{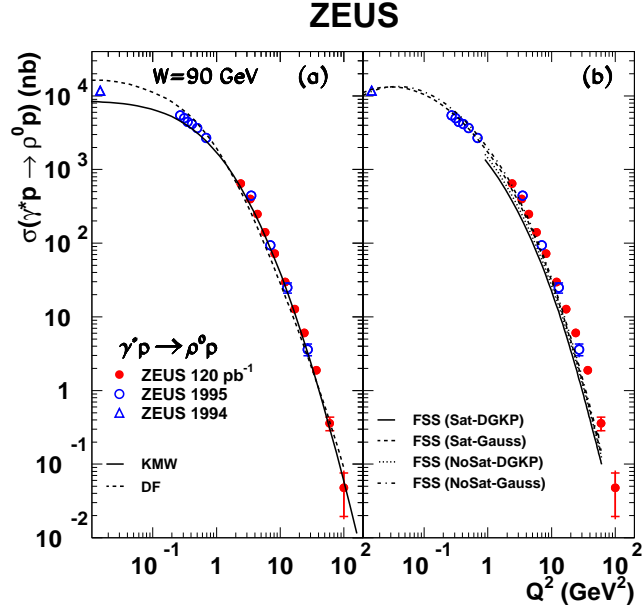}}
\caption {The $\gamma^*p\rightarrow\rho^0p$ cross section as a function of $Q^2$ measured at $W=90\;GeV^2$ and comared in (a) and (b) with different models as described in the text.}
\label{q2_rho}
\end{figure}

Recently, a precision measurement of the reaction $\gamma^*p\rightarrow\rho^0 p$ was published by ZEUS~\cite{zeus_rho}. It was found that the cross section falls steeply with the increasing of $Q^2$ but, unlike it was observed for the $J/\psi$ electroproduction~\cite{zeus_jpsi,h1_jpsi}, it cannot be described by a simple propagator term like $\sigma\propto (Q^2+M^2)^{-n}$, in particular an $n$ value increasing with $Q^2$ appears to be favored. Figure~\ref{q2_rho} reports the cross section for the $\rho^0$ electroproduction versus $Q^2$ compared with several theoretical predictions: the KWM model~\cite{KMW} based on the saturation model, the FSS model~\cite{FSS} with and without saturation and the DF model~\cite{DF}. 

%%\subsection{The $W$ dependence}

\begin{figure}[htbp]
%\epsfxsize=10cm   %width of figure - will enlarge/reduce the figures
%\epsfbox{fig3.eps}
%\figurebox{2cm}{3cm}{} %to have a box alone 
\centering
\includegraphics[width=0.55\textwidth,angle=0]{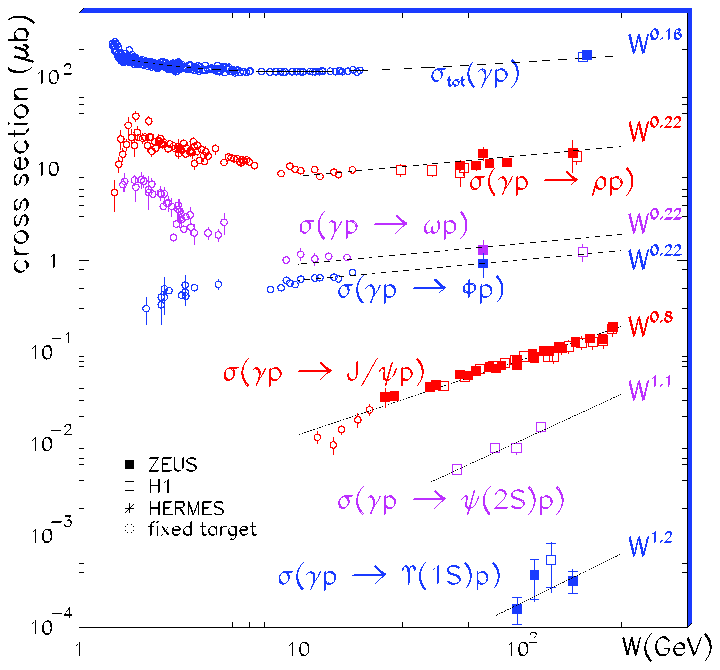}
%%\centerline{\epsfxsize=4.1in\epsfbox{W_php.eps}}   
\caption{The $W$ dependence of the cross section for exclusive VM photoproduction together with the total photoproduction cross section. Lines are the result of a $W^{\delta}$ fit to the data at high $W$-energy values. \label{W_php}}
\end{figure}

The soft to hard transition can be observed looking at the W-dependence of the VMs photoproduction ($Q^2=0$), where the scale is provided by $M^2$. Figure~\ref{W_php} collects the $\sigma ( \gamma^* p\rightarrow V p )$ as a function of $W$ from the lightest vector meson, $\rho^0$, to the heaviest, $\Upsilon$, compared to the total cross section. 

The cross section rises with the energy as $W^{\delta}$, where the $\delta$ exponent increases with the hard scale $M^2$ as expected for a transition from the soft to the hard regime. New results on the $\Upsilon$ photoproduction~\cite{upsilon}, recently published by ZEUS, confirmed the steeper rise of $\sigma(W)$ for higher vector meson masses. 

\begin{figure}[htbp]
%\epsfxsize=10cm   %width of figure - will enlarge/reduce the figures
%\epsfbox{fig3.eps}
%\figurebox{2cm}{3cm}{} %to have a box alone 
%%\centering
\begin{tabular}{lr}
%%\begin{tabular}{c}
%%\includegraphics[width=0.5\textwidth,angle=0]{d09-109f3a.eps}\\
\includegraphics[width=0.5\textwidth,angle=0]{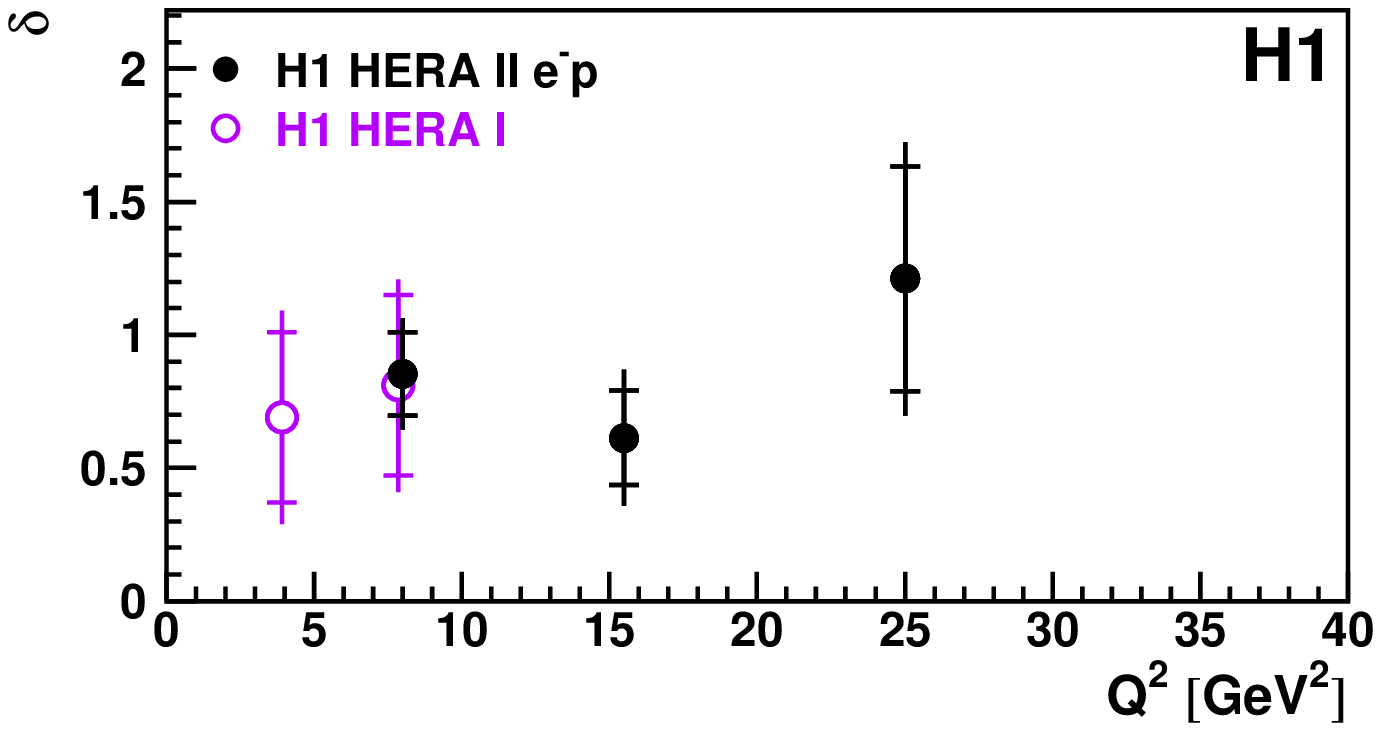} &
%%\end{tabular}  
%%\includegraphics[width=0.5\textwidth,angle=0]{DESY-08-132_3.eps}
%%\includegraphics[width=0.5\textwidth,angle=0]{d07-142f3b.eps}A.B. Kaidalov, V.A. Khoze, A.D. Martin, M.G. Ryskin   
\includegraphics[width=0.5\textwidth,angle=0]{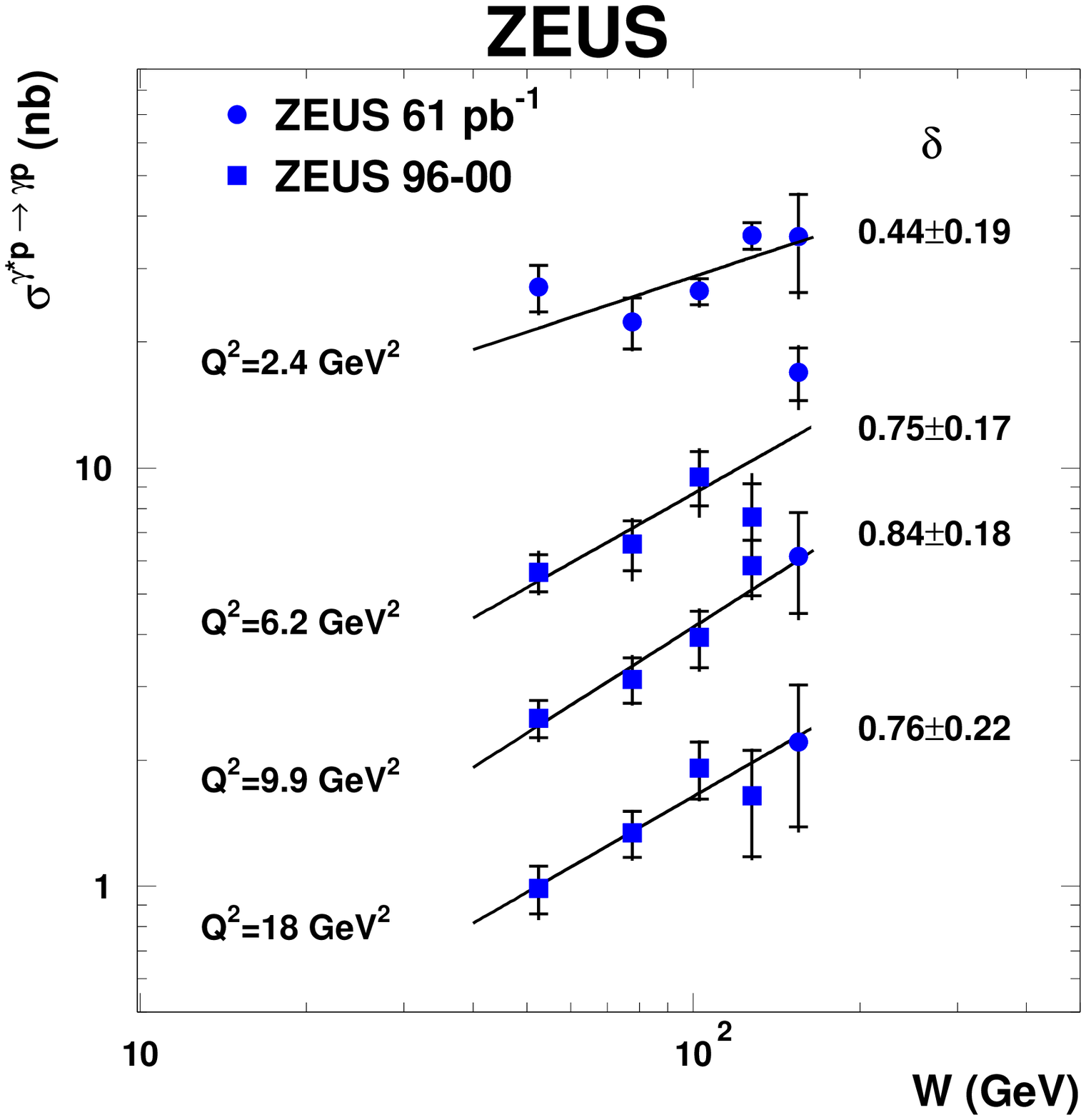}
\end{tabular}  
%%%%\includegraphics[width=0.8\textwidth,angle=0]{W_dvcs_collec.eps}
%%\centerline{\epsfxsize=4.1in\epsfbox{W_dvcs_collec.eps}}   
\caption{The $W$ dependence of the cross section for a DVCS process. Lines come from a $W^{\delta}$ fit to the data. Left: the H1 measurement of the $\delta$ slope as a function of $Q^2$. Right: the new ZEUS measurement at low $Q^2$ (dots) together with the published measurements (squares). \label{W_dvcs}}
\end{figure}

The transition from the soft to the hard regime can also be studied varying $Q^2$. Recent results were achieved by H1~\cite{h1_dvcs} and ZEUS~\cite{zeus_dvcs} for the exclusive production of a real photon, the Deeply Virtual Compton Scattering (DVCS), where the hard scale is provided only by the photon virtuality, $Q^2$. Figure~\ref{W_dvcs} shows the H1 (left) and the ZEUS (right) results. A similar result was obtained for the $J/\psi$ electroproduction~\cite{zeus_jpsi,h1_jpsi}.
 
\begin{figure}[htbp]
%\epsfxsize=10cm   %width of figure - will enlarge/reduce the figures
%\epsfbox{fig3.eps}
%\figurebox{2cm}{3cm}{} %to have a box alone 
\centering
\begin{tabular}{cc}
\includegraphics[width=0.53\textwidth,angle=0]{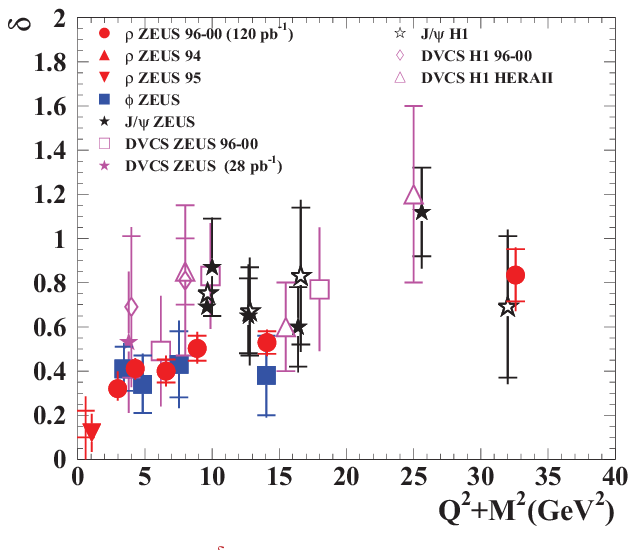}
%%\centerline{\epsfxsize=4.1in\epsfbox{W_php.eps}}
\includegraphics[width=0.46\textwidth,angle=0]{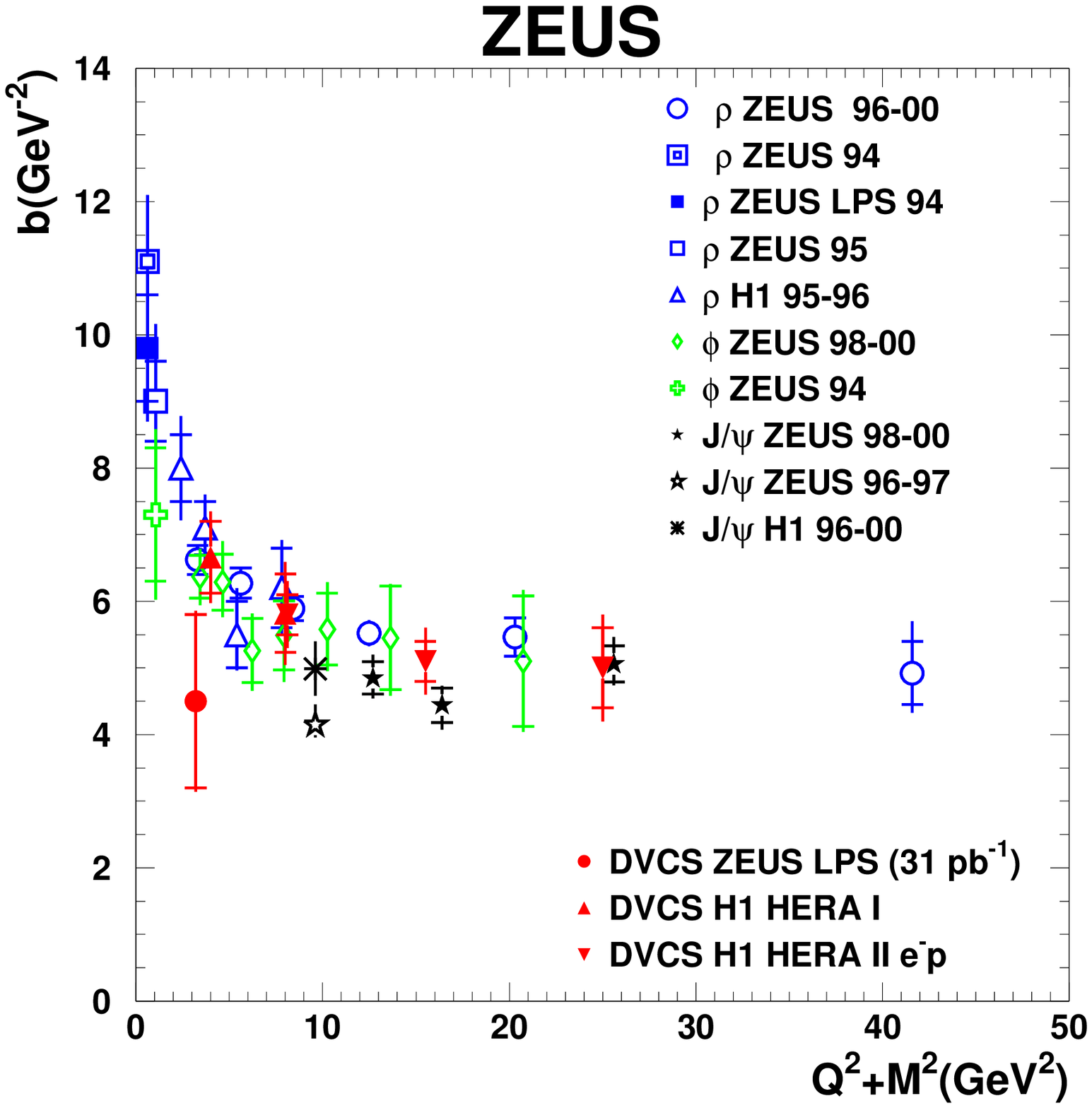}
\end{tabular}   
\caption{The dependence on the hard scale $Q^2+M^2$ of the value $\delta$ (left) extracted from a fit $W^{\delta}$ and of the slope $B$ ($b$ in the figure lable) (right) extracted from a fit $\frac{d\sigma}{dt}\propto e^{B|t|}$ for the exclusive VM electroproduction. DVCS is also included. \label{W_dis}}
\end{figure}

The electroproduction of a large variety of VMs was studied at different $Q^2$ values and the corresponding slope $\delta$ is reported in Fig.~\ref{W_dis} (left) versus the scale $Q^2+M^2$, including the DVCS measurements. 
The data behaviour seems to be universal with $\delta$ rising from 0.2, as expected from a soft Pomeron exchange~\cite{Wsoft}, showing a logarithmic shape $\delta\propto \ln(Q^2+M^2)$.

The steep rise with $W$ of the cross section even at low-$Q^2$, seems to suggest that the most sensitive part to the soft scale comes from the wave function of the pruduced VM.

\subsection{$t$ dependence of the cross section and GPDs}

The differential cross section as a function of $t$, can be parametrised by an exponential fit: $\frac{d\sigma}{d|t|}\propto e^{b|t|}$. Figure~\ref{W_dis} (right) reports the collection of the $b$ values versus the scale $Q^2+M^2$ for the electroproduction of VMs and DVCS, with $b$ decreasing from $\sim 11\; GeV^{-2}$ to $\sim 5\; GeV^{-2}$ as expected in hard regime. 
Since the $b$ value can be related via a Fourier transform to the impact parameter and assuming that the exclusive process in the hard regime is dominated by gluons, the relation $\langle r^2\rangle=2b(\hbar c)^2$ can be used to obtain the radius of the gluon confinement area in the proton. $b\sim 5\;GeV^2$ corresponds to $\langle r^2\rangle\sim 0.6\;fm$ smaller than the proton radius ($\sim 0.8\;fm$) indicating that the gluons are well contained within the charge-radius of the proton.

Measurements of the $t$-slope parameters $b$ are key measurements
for almost all exclusive processes. Indeed, a Fourier
transform from momentum to impact parameter space
readily shows that the $t$-slope is related to the typical
transverse distance between the colliding objects. At
high scale, the \={q}q dipole is almost point-like, and the $t$
dependence of the cross section is given by the transverse
extension of the gluons (or sea quarks) in the proton for
a given $x_{Bj}$ range. In particular for DVCS, interpretation
of t-slope measurements does not suffer from
the lack of knowledge of the VM wave function.
Then, a DVCS cross section, differential in $t$, is directly related to
GPDs~\cite{b_slope_papers}.

\begin{figure}[htbp]
\centering
\begin{tabular}{cc}
\includegraphics[width=0.5\textwidth,angle=0]{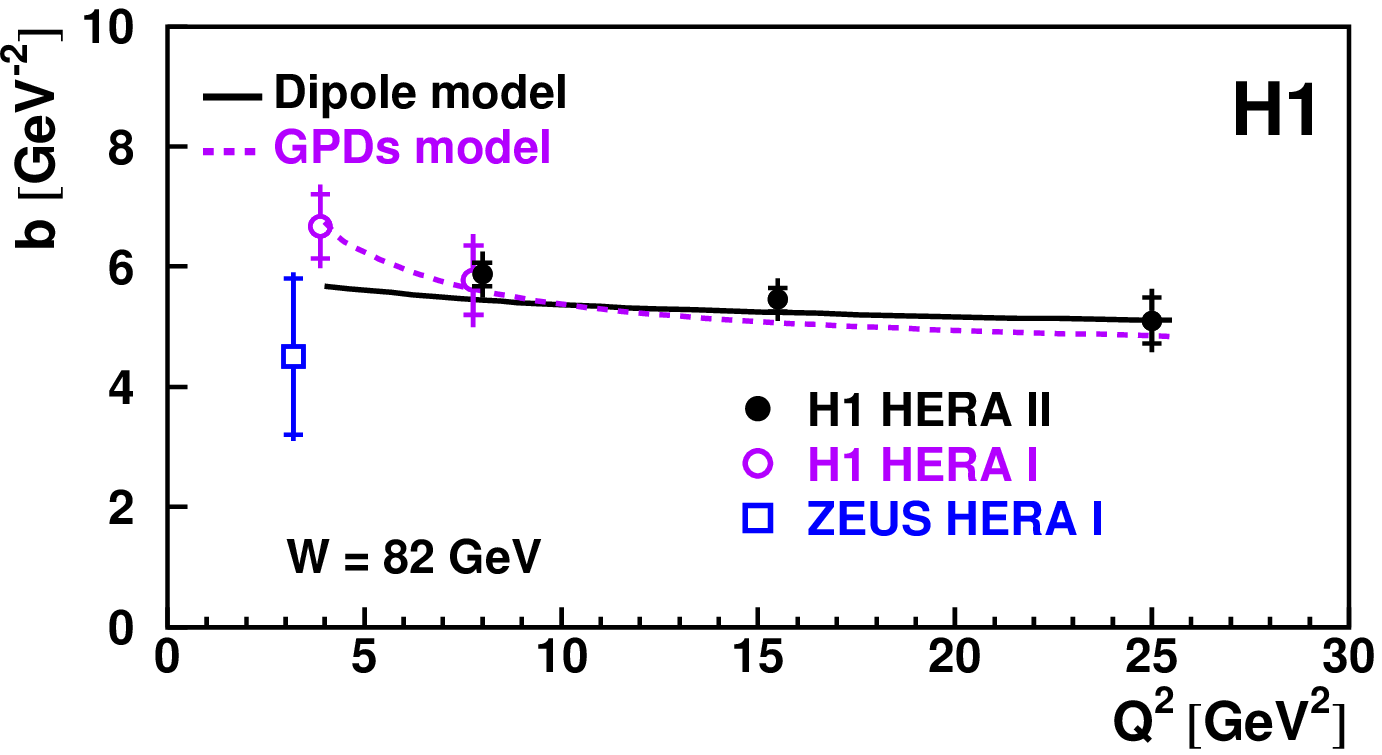}
\includegraphics[width=0.5\textwidth,angle=0]{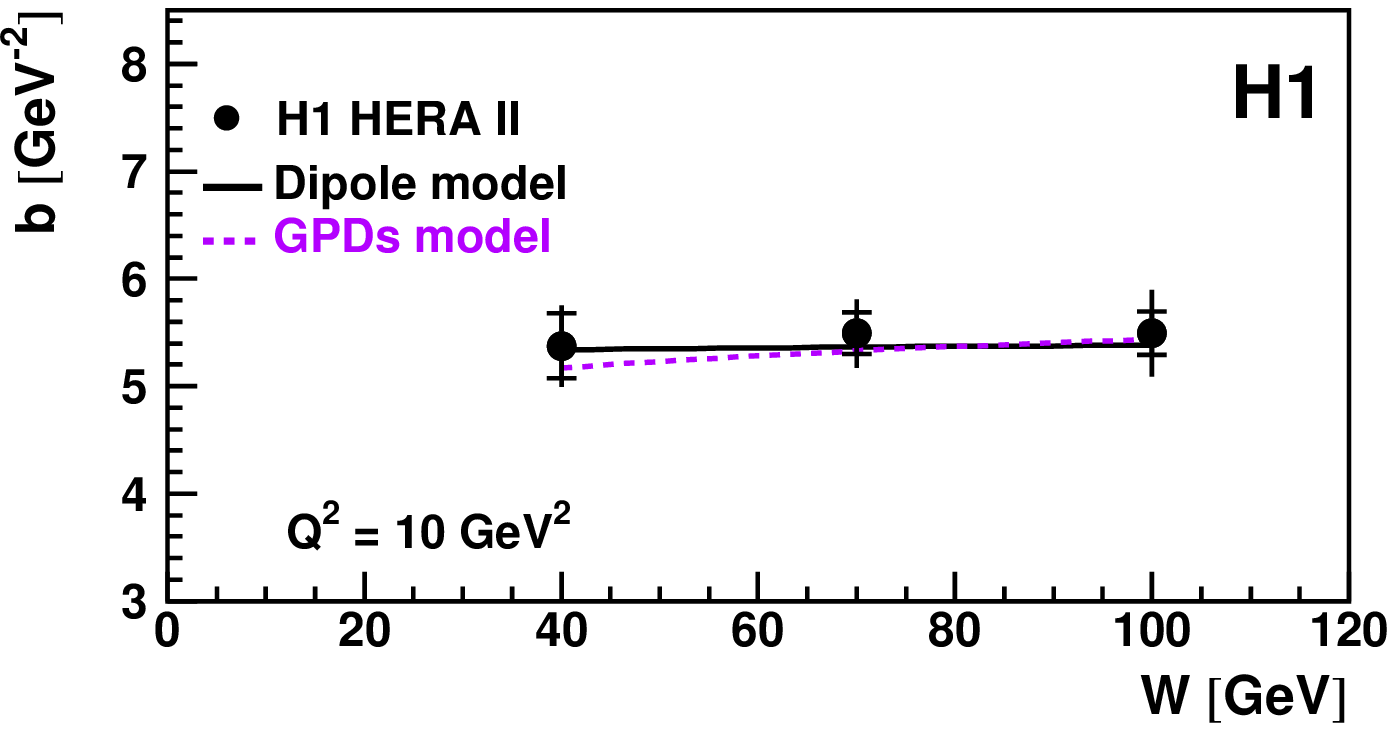}
\end{tabular}
\caption{The $t$ slope parameter, $b$, as a function of $Q^2$ (left) and $W$ (right).}
\label{b_dvcs}
\end{figure}

The measurement of $d\sigma/d|t|$ for the DVCS process, recentrly published by the H1 Collab~\cite{h1_dvcs}, where $t$ was obtained from the transverse momentum distribution of the photon, studied $b$ versus $Q^2$ and $W$ as shown in Fig.~\ref{b_dvcs}. $b$ seems to decrease with $Q^2$ up to the value expected for a hard process but it doesn't depend on $W$. 
  
A new ZEUS measurement~\cite{zeus_dvcs} of $d\sigma/d|t|$ has been achieved from a direct measurement of the proton final state of using a spectrometer based on the roman pot thechnique (see Fig.~\ref{b_dvcs},right). The result $b=4.5\pm 1.3~(stat.)\pm 0.4~(syst.)~GeV^{-2}$, measured at $Q^2=3.2~GeV^2$ and $W=104~GeV$, is consistent, within the large uncertainties due to the low acceptance of the spectrometer, with the H1 result~\cite{h1_dvcs} of $b=5.45\pm 0.19~(stat.)\pm 0.34~(syst.)~GeV^{-2}$ at $Q^2=8~GeV^2$ and $W=82~GeV$.

\begin{figure}[htbp]
\centering
\begin{tabular}{cc}
\includegraphics[width=0.5\textwidth,angle=0]{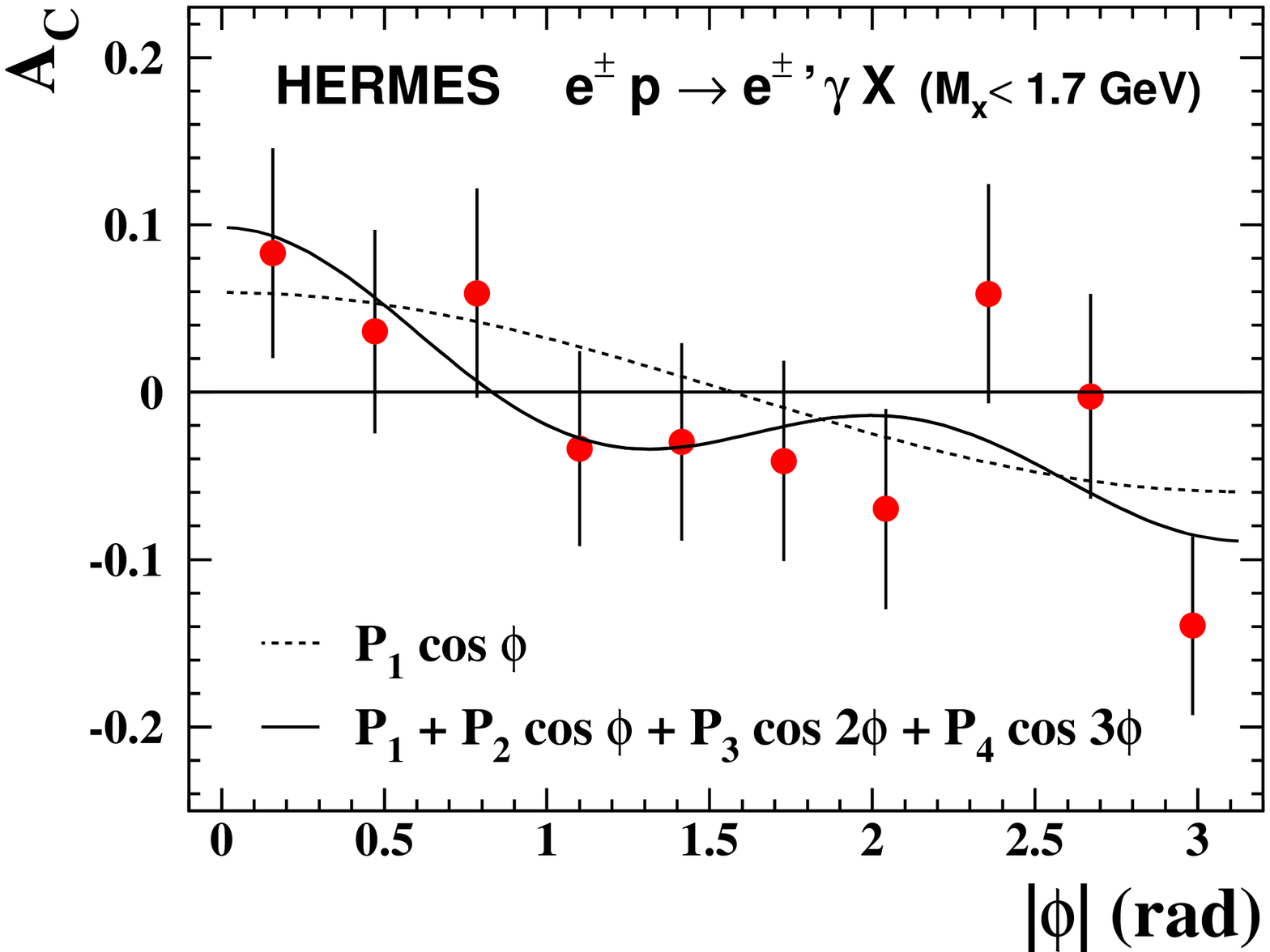}
\includegraphics[width=0.5\textwidth,angle=0]{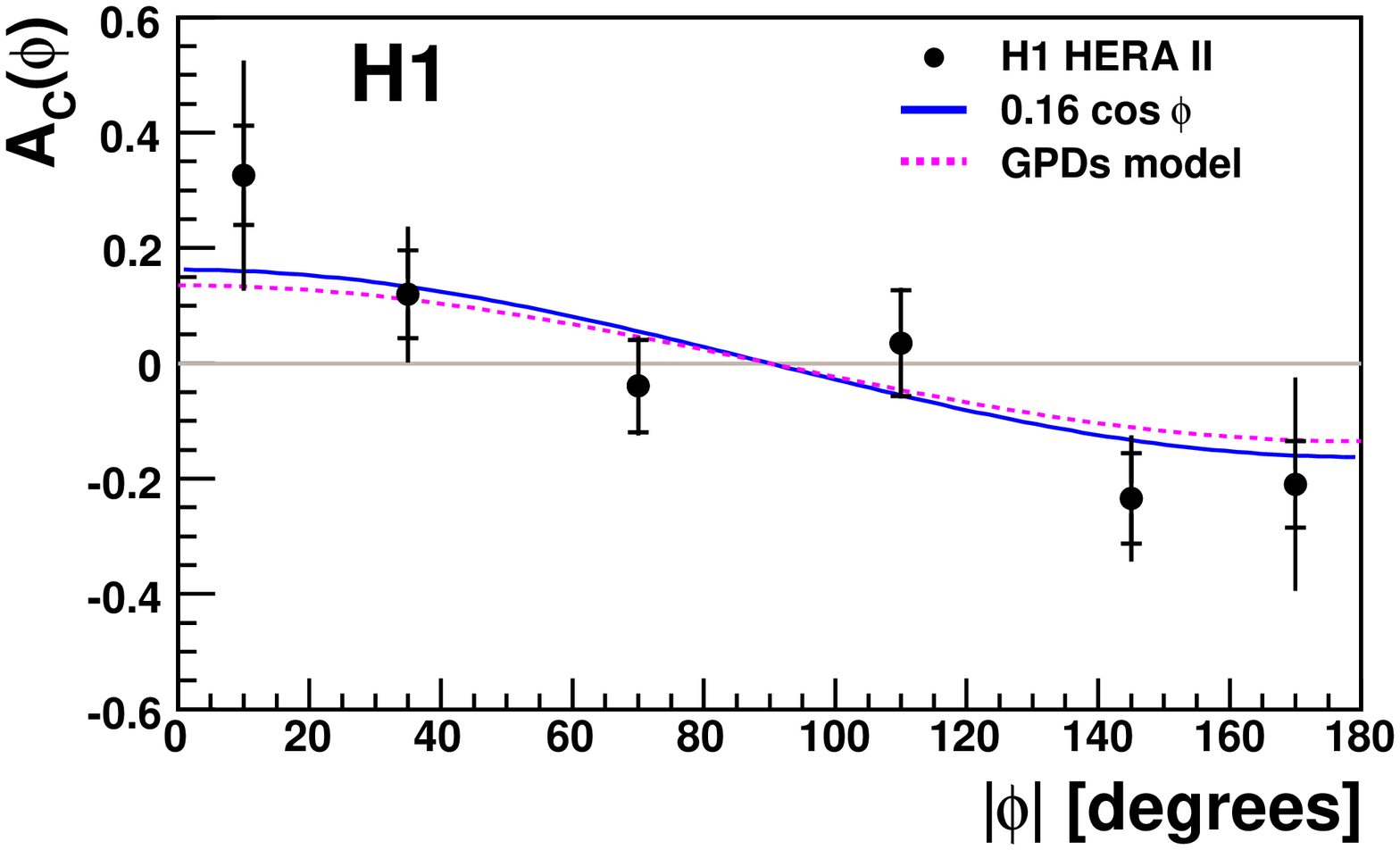}
\end{tabular}
\caption{The BCA as a function of the azimuthal angle between the production and the scattering plane, measured by the HERMES (left) and H1 (right) experiments at HERA.}
\label{BCA}
\end{figure}

The complete parton imaging in the nucleon would
need to get measurements of $b$ for wide range of $x_{Bj}$ values, $0.001 < x_{Bj} > 0.1$, that experimentally appears to be really difficult. In fact, there is one
way to recover $x_{Bj}$ and $t$ correlations over the whole $x_{Bj}$
domain: to measure a Beam Charge Asymmetry (BCA).
A determination of a cross section asymmetry with
respect to the beam charge has been realized by the H1~\cite{h1_dvcs} and HERMES~\cite{hermes_dvcs}
experiments by measuring the ratio $(d\sigma^+ - d\sigma^-) / (d\sigma^+ + d\sigma^-)$ as a function of the azimuthal angle, $\phi$, between the production and the scattering plane. The results are shown in Fig.~\ref{BCA}.

\section{Exclusive diffraction at LHC}

Interest in diffraction at the LHC has been greatly boosted recently by theoretical predictions~\cite{KMR} that identified central exclusive production (CEP) as a potential discovery channel for the Higgs boson. In the last years, both ATLAS and CMS have set up forward physics programs. Both experiments have Zero Degree Calorimeters with acceptance for neutral particles for $|\eta| > 8.3$. Apart from those, their forward instrumentation is quite complementary. ATLAS is equipped with a dedicated luminosity system
consisting of ALFA, Roman-pot housed tracking detectors at 240 m from the interaction point (I.P.), which will peform an absolute luminosity measurement in runs with special LHC optics, and of LUCID, Cherenkov detectors ($5.6 < |\eta| < 6.0$) for the primary purpose of luminosity monitoring during routine LHC data taking. At the CMS I.P., the task of an absolute luminosity determination will be carried out by an independent experiment, TOTEM, with Roman-pot housed silicon detectors at 220 m distance from the I.P. and two tracking
telescopes inside of the CMS volume. CMS in addition has the CASTOR calorimeter which extends the CMS calorimetric coverage to rapidity values of 6.5. CASTOR gives access to the QCD parton evolution dynamics at very low-$x$.

In addition, FP420, a joint R\&D program of ATLAS, CMS and the LHC machine group has investigated the
feasibility of an upgrade of the forward detector instrumentation to make possible the direct observation of the scattered protons in CEP of a Higgs boson.

\section{The Pomeron at HERA, a two-pole model}
A simple factorized Regge-pole model~\cite{DVCS} for the description of the DVCS amplitude was suggested and
successfully tested over the HERA data. 
The authors are now working to include in the analysis the
VMs production by using and extending the
main ideas of the model.

It follows from perturbative QCD that asymptotically the Pomeron
is an infinite sum of poles. This result is far from practical
realization, however in minimal version, it is legitimate to
assume that the Pomeron is a sum of two Regge poles
\begin{equation}\label{TwoP}
A(s,t,\tilde Q^2)=f_s(s,t,\tilde
Q^2)(-is/s_0)^{\alpha_s(t)}+f_h(s,t,\tilde
Q^2)(-is/s_0)^{\alpha_h(t)},
\end{equation}
where the subscripts allude that the first term is ``soft", with
$\alpha_s(0)\approx 1.08$ and the second one is ``hard", with
$\alpha_s(0)\approx 1.4,$ these numbers coming from the fits to
soft hadronic and hard deep inelastic reactions. Such a
two-component Pomeron was first suggested by Landshoff \cite{L} and it
accounts for both soft and hard processes in the framework of
a universal $Q^2$-independent Pomeron, which implies that the
trajectories $\alpha_s$ and $\alpha_h$ are the same in all
reactions, differing only by their weights, determined by the
residue functions $f_i(s,t,\tilde Q^2),\ \ i=1,2$. 
According to arguments~\cite{Kaidalov} based on
unitarity, $f_h$ is progressively damped more that $f_s$ with
increasing $Q^2+M^2$ hence the hard term is masked at low
$Q^2$ (soft reactions) while it dominates at high
$Q^2$. 

%\newpage
%\vspace{1cm}
\begin{figure}[h]
\begin{center}
\includegraphics[clip,scale=0.7]{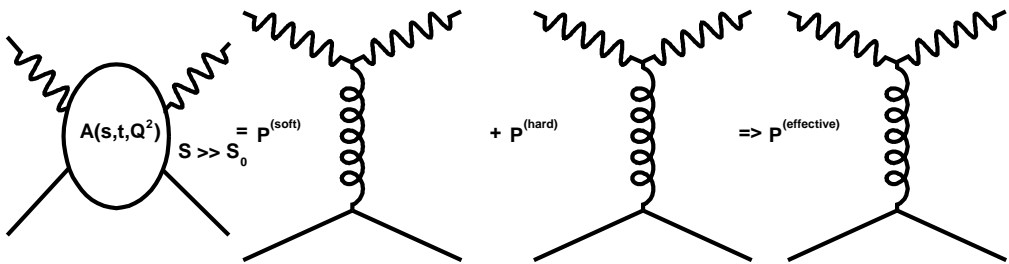}
\end{center}
%\vspace{-10cm}
 \caption{\small\it{Sum of two Regge-Pomeron exchanges,
approcimated by an "effective Pomeron" (rightmost diagram). }}
\label{fig:diagram2}
\end{figure}

In the model the two (soft and hard) terms have identical
functional form, differing only by the parameters of two trajectories,
$\alpha_s(t)$ and $\alpha_h(t)$. This is
a universal Pomeron in the sense, that its trajectories are
the same in any (soft or hard) process, varying only the
relative weight of the two terms
\begin{eqnarray}\label{2}
A(s,t,Q^2)=A^s+h A^h=-[e^{b_1\alpha^s(t)}e^{b^s_2
\beta^s(z)}(-is/s_0)^{\alpha^s(t)}+\nonumber \\ 
 h e^{b_1\alpha^h(t)}e^{b^h_2
\beta^h(z)}(-is/s_0)^{\alpha^h(t)}].
\end{eqnarray}

%%\section{Section 1}

%%Please, mimimize the number of extra $LATEX$ packages.
%%Formulas as well as figures should fit the page.

%%\begin{equation}
%%L \propto M
%%\end{equation}
%%\begin{eqnarray}
%%L \propto M\\
%%L \propto M\\
%%L \propto M
%%\end{eqnarray}

%%\section {Section 2}

%%If you don't know how to merge pictures into one figure,
%%just put them separatly with own captions.

%%\begin{figure}[t]
%%{\includegraphics[width=6.5cm]{123.eps}}
%%\caption {Typical background event}
%%\end{figure}

\end{document}